\documentclass[10pt,journal,comsoc]{IEEEtran}
\usepackage{graphicx}
\usepackage{amsmath,amssymb}
\usepackage{booktabs}
\usepackage{cite}
\usepackage{hyperref}

\hypersetup{colorlinks=true,
	linkcolor=blue,
	citecolor=blue,
	urlcolor=black,
}

\title{B2X Networks: Joint Design of Communication and Control for Embodied Intelligence}

\author{Yuanwei Liu~\IEEEmembership{Fellow,~IEEE}, Xu Gan, Zhaolin Wang, Chongjun Ouyang, Hao Jiang, Zongyao Zhao,\\ Kaibin Huang~\IEEEmembership{Fellow,~IEEE}, and Robert Schober~\IEEEmembership{Fellow,~IEEE}

\thanks{Y. Liu, X. Gan, Z. Wang, Z. Zhao, and K. Huang are with the Department of Electrical and Computer Engineering, The University of Hong Kong, Hong Kong (e-mails: \{yuanwei, eee.ganxu, zhaolin.wang, zongyao, huangkb\}@hku.hk).}
\thanks{C. Ouyang and H. Jiang are with the School of Electronic Engineering and Computer Science, Queen Mary University of London, London E1 4NS, U.K. (e-mails: \{c.ouyang, hao.jiang\}@qmul.ac.uk).}
\thanks{R. Schober is with the Institute for Digital Communications, Friedrich-Alexander-Universität Erlangen-Nürnberg (FAU), 91054 Erlangen, Germany. (e-mail: robert.schober@fau.de).}
}

\begin{document}
\maketitle

\begin{abstract}
This article proposes the concept of \emph{brain-body-to-everything (B2X)} networks to facilitate the integration of wireless networks and embodied intelligence. In this framework, the \emph{brain} refers to the intelligence functions for reasoning, planning, and decision-making, the \emph{body} denotes the physical embodied agent that senses and acts in the real world, and \emph{X} represents the surrounding ecosystem involved in the brain-body interaction loop. Two B2X architectures with \emph{distributed} and \emph{centralized} brains are introduced to characterize different placements of intelligence across the body, base station, and core network. The uplink and downlink designs of B2X networks are then discussed under a representative base-station-side brain setting. For the uplink, communication is redesigned for B2X state acquisition under event urgency, sensing volume, and simultaneous multi-body access. For the downlink, communication is redesigned to coordinate command delivery and conventional service under shared radio resources. Based on these uplink and downlink considerations, a communication-control Pareto boundary is further used to characterize the loop-level trade-off between wireless transmission performance and control quality in B2X networks. Finally, several open research problems are discussed to guide future B2X network design.
\end{abstract}

\section{Introduction}
Embodied intelligence is transforming artificial intelligence (AI) from a digital capability to a physical, interactive, and adaptive form of intelligence~\cite{duan2022survey}. This trend is also reflected in the recent rise of physical AI, where AI systems are expected to perceive, reason, and act through physical bodies in real-world environments~\cite{agarwal2025cosmos}. Unlike conventional AI systems that mainly process static cloud data, embodied intelligent agents, including robots, autonomous vehicles, drones, extended-reality devices, and intelligent sensors, need to sense the physical world, reason under uncertainty, and act in real time~\cite{xu2025embodied}. Their intelligence is tied to the operating environment and depends directly on wireless networks. In particular, they need to exchange task states, control commands, model updates, and contextual knowledge with peer agents and network-native intelligence. Without reliable, low-latency, and context-aware wireless support, their operation is constrained by the computation, energy, sensing range, and memory available at individual devices~\cite{zhang2025comai,zhang2026embodied}. 

Edge intelligence provides a natural bridge between embodied agents and wireless infrastructure~\cite{zhou2019edge}. By enabling computation, learning, inference, and decision-making at radio access networks (RANs), wireless systems can facilitate information transfer and edge-side intelligence processing~\cite{letaief2021edge}. Such networks assist embodied intelligence through low-latency inference and control, shared environment representations, and site-specific knowledge for decisions beyond the local sensing range. In this view, the wireless network becomes both a communication medium and a distributed nervous system that connects distributed intelligence with physical-world operation.

The current technological landscape presents an opportune moment to investigate the convergence of embodied intelligence and wireless networks. AI-RAN has recently attracted attention from academia and industry~\cite{khan2023airan,kundu2026airan}. The AI-RAN Alliance coordinates industry and academia on AI-powered mobile networks~\cite{kundu2026airan,polese2026beyond}, while the open radio access network (O-RAN) ecosystem~\cite{alam2025oran} promotes open, virtualized, and intelligent RAN architectures with AI-enabled management. These developments indicate a broader transition toward wireless networks that support human-centric services and machine-centric intelligence services involving sensing, learning, reasoning, and control. Embodied intelligence motivates AI-RAN development, while AI-RAN provides the infrastructure needed for scalable and responsive embodied intelligence.

However, supporting embodied intelligence over wireless networks requires more than higher data rates or lower latency alone. Embodied agents operate in closed loops by sensing the environment, communicating relevant information, receiving decisions or commands, and acting on the physical world. Communication errors, latency, packet loss, and resource constraints directly affect control performance, safety, and task success, while the control objective determines the content, timing, and accuracy requirements of transmission. This makes separate communication and control design increasingly inadequate. Joint design of communication and control (JDCC)~\cite{gan2026modeling} is therefore central to integrating embodied intelligence with wireless networks. In the uplink, the network needs to acquire timely and task-relevant state information from embodied agents, including sensory observations, local decisions, and event-triggered feedback. In the downlink, the network needs to deliver control commands, model updates, coordination signals, or task instructions with reliability and timeliness matched to the physical dynamics. Rather than treating all bits equally, future networks need to account for the value of information for control. For example, stale but accurate data may be less useful than timely coarse information, and high-rate sensing streams may be unnecessary if compact task-oriented features are sufficient for decision making.

In this article, we propose the \emph{brain-body-to-everything (B2X)} network for integrating embodied intelligence with wireless systems. B2X treats the wireless infrastructure as part of the embodied intelligence loop, in which the \emph{body} senses and acts in the physical world, the \emph{brain} performs reasoning and decision-making, and \emph{X} provides external agents, infrastructure, and task context. We define the B2X concept and introduce distributed-brain and centralized-brain architectures according to intelligence placement. We then focus on a representative base-station (BS)-side centralized brain architecture and discuss uplink state acquisition and downlink command delivery as two wireless interfaces for embodied operations. We further use a Pareto-boundary characterization to capture the loop-level communication-control trade-off in B2X operation. Finally, the article identifies open problems regarding loop-level metrics, deployment, function- and energy-aware protocols, theory, and scalable distributed intelligence.

\begin{figure*}[!t]
\centering
\includegraphics[width=0.55\textwidth]{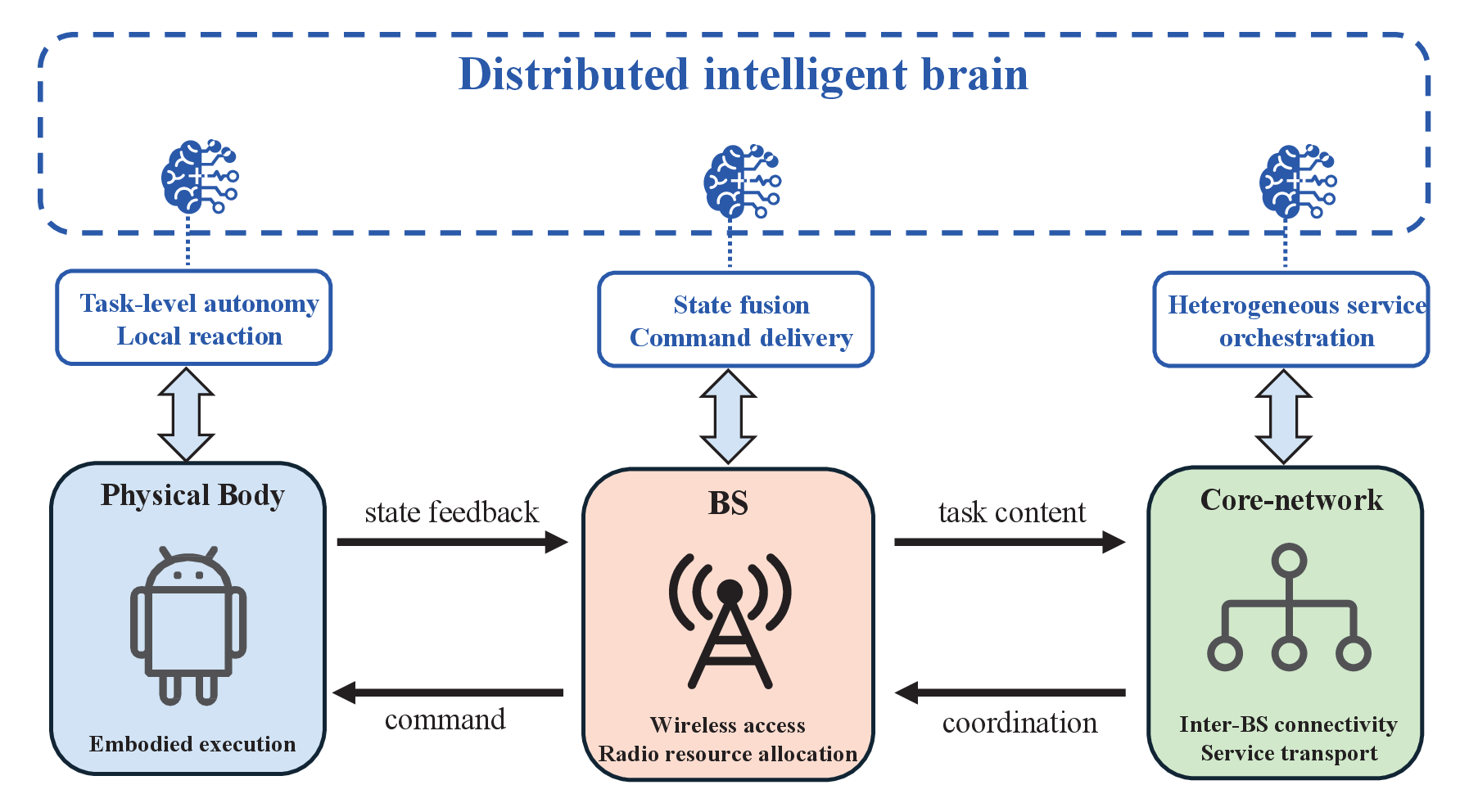}
\caption{Distributed B2X brain placement across body-side, BS-side, and CN-side domains.}
\label{fig:brain_architecture}
\end{figure*}

\section{Fundamentals of B2X Networks}

This section introduces the basic concept of B2X networks. B2X provides a network-level framework for supporting embodied intelligence over wireless infrastructure. Unlike conventional wireless systems that mainly focus on information delivery, B2X networks emphasize closed-loop interaction among sensing, communication, computing, and control. In this framework, wireless networks are not only connectivity providers. They also serve as intelligence-bearing infrastructure that assists embodied agents in understanding, coordinating, and acting in the physical world.

\subsection{What is B2X?}

B2X refers to a wireless-network-supported architecture for embodied intelligence, where the intelligence, the physical executor, and the external ecosystem are connected via a closed-loop interaction. A B2X network contains three essential components: \emph{brain}, \emph{body}, and \emph{X}.
\begin{itemize}
\item The \emph{brain} refers to the intelligence that supports reasoning, planning, and decision-making. In B2X networks, the brain may be placed at the wireless edge, distributed across network nodes, or partly retained inside the body. This network-side brain gives the embodied agent access to stronger computation and broader contextual knowledge than local intelligence alone.
\item The \emph{body} refers to the physical executor that interacts with the real world. It observes the environment, receives decisions or commands from the brain, and converts them into physical actions. The body is the physical carrier of embodied intelligence, since it turns digital decisions into real-world actions.
\item The \emph{X} represents the external ecosystem connected to the brain-body loop. It includes the environment, surrounding agents, infrastructure, and task context that influence the embodied operation. X emphasizes that the brain and body continuously adapt to external conditions and benefit from the information and coordination provided by the surrounding ecosystem.

\end{itemize}

Based on this structure, B2X networks operate in a closed loop. The body observes the physical world and reports task-relevant information to the brain through the wireless network. The brain interprets this information and combines it with external knowledge from X to generate decisions, which are delivered to the body and translated into physical actions. These actions change the environment, and the body senses the new state to continue the loop. This loop relies on sensing, communication, computing, and control. \emph{Sensing} captures body and environmental states. \emph{Communication} exchanges observations, decisions, and coordination information among the body, brain, and X. \emph{Computing} maps information into predictions, decisions, and plans. \emph{Control} converts decisions into physical actions and guides body-environment interaction. These functions are tightly coupled because sensing determines the available decision information, communication determines exchange timeliness and reliability, computing determines information interpretation, and control determines physical-world effects. B2X networks should therefore not be treated as a simple combination of independent modules. They require a unified design for the entire embodied loop.

The B2X concept clarifies the roles of intelligence, physical execution, and external ecosystem. It also highlights the functional coupling of the embodied loop. Based on the placement of the brain, the following subsections discuss \emph{distributed-brain B2X networks} and \emph{centralized-brain B2X networks}.

\subsection{B2X with Distributed Brains}

In the distributed-brain architecture, B2X intelligence is not assigned to one location. Instead, it is partitioned across the body-side, BS-side, and core-network (CN)-side brains according to the time scale and spatial scope of the embodied task. As illustrated in Fig.~\ref{fig:brain_architecture}, the body-side brain supports short-window reaction, local safety, and basic autonomy when wireless assistance is delayed or unavailable. The BS-side brain fuses time-sensitive state reports, interprets local context, and coordinates command delivery at the radio-access edge, while the CN-side brain provides a broader cross-BS view for policy coordination, continuity, and long-horizon orchestration. This partitioning reflects the multiple time scales of embodied operation. Local response, BS-side physical and channel knowledge, and CN-side cross-BS visibility are therefore suited to different functions, including safety reaction, cooperative sensing, radio-aware command scheduling, mobility management, service migration, and model or policy synchronization. Distributed B2X makes brain placement a control-design variable for assigning fresh information, decision authority, and wireless-resource coordination to suitable locations for physical execution.

The distributed view in Fig.~\ref{fig:brain_architecture} highlights the main advantage of B2X brain placement: embodied intelligence is not constrained by a single latency and reliability regime. Event-triggered state updates, over-the-air aggregation, control-aware scheduling, beam management, and brain-state transfer further help the distributed brains maintain a common embodied loop. At the same time, distributed brains introduce coupling beyond conventional communication services. The system needs to determine authority, maintain consistency between local and network-side decisions, and detect stale task states before they degrade control. In B2X networks, the benefits of distributed intelligence can be fully realized when different brains share the state, transfer authority, and maintain control consistency under wireless delay, mobility, and resource constraints.

\subsection{B2X with a Centralized Brain}

A centralized B2X architecture provides the complementary view. In this case, the brain is concentrated at one dominant side of the system, while the other sides mainly provide sensing, wireless connectivity, physical execution, or ecosystem support. Centralization does not imply reduced intelligence. It is useful when one decision point naturally dominates the closed-loop task or when reduced coordination complexity is more important than fully distributed intelligence. The suitable placement depends on reaction time, external context, and the spatial range of task coordination.

\begin{itemize}
    \item \textbf{Body-side centralized brain:} The body-side brain carries the dominant intelligence when the task is governed by hard real-time local control. Highway autonomous driving and industrial robotic-arm safety stopping are representative examples, since sensing and actuation need to continue under imperfect network assistance.
    \item \textbf{BS-side centralized brain:} The BS-side brain becomes the dominant intelligence point when local observations are insufficient and the embodied operation depends on wireless network information, local state fusion, and coordinated command delivery. Urban-intersection driving and smart-factory multi-robot collaboration are typical cases, since multiple bodies share the physical environment and wireless resource pool.
    \item \textbf{CN-side centralized brain:} The CN-side brain becomes the dominant intelligence point when the task spans multiple BSs, regions, or service domains. City-scale vehicular coordination and large-scale unmanned aerial vehicle (UAV) supervision emphasize regional planning, policy consistency, and service continuity rather than millisecond-level command generation for every action.
\end{itemize}

Among these placements, the BS-side centralized brain is particularly useful for studying B2X over wireless links. It is close enough to the physical bodies to affect sensing freshness and command latency, while providing computing and resource-control capability for network-side reasoning, task-oriented state fusion, and radio resource control. This position exposes the key tension of B2X. The BS serves as both a communication node and an intelligence node, so its resource decisions affect throughput, reliability, state-estimation quality, command timing, and closed-loop control. For this reason, the remainder of this article uses the BS-side brain as a representative architecture for discussing uplink and downlink B2X design.

\begin{figure*}[t]
\centering
\includegraphics[width=0.7\textwidth]{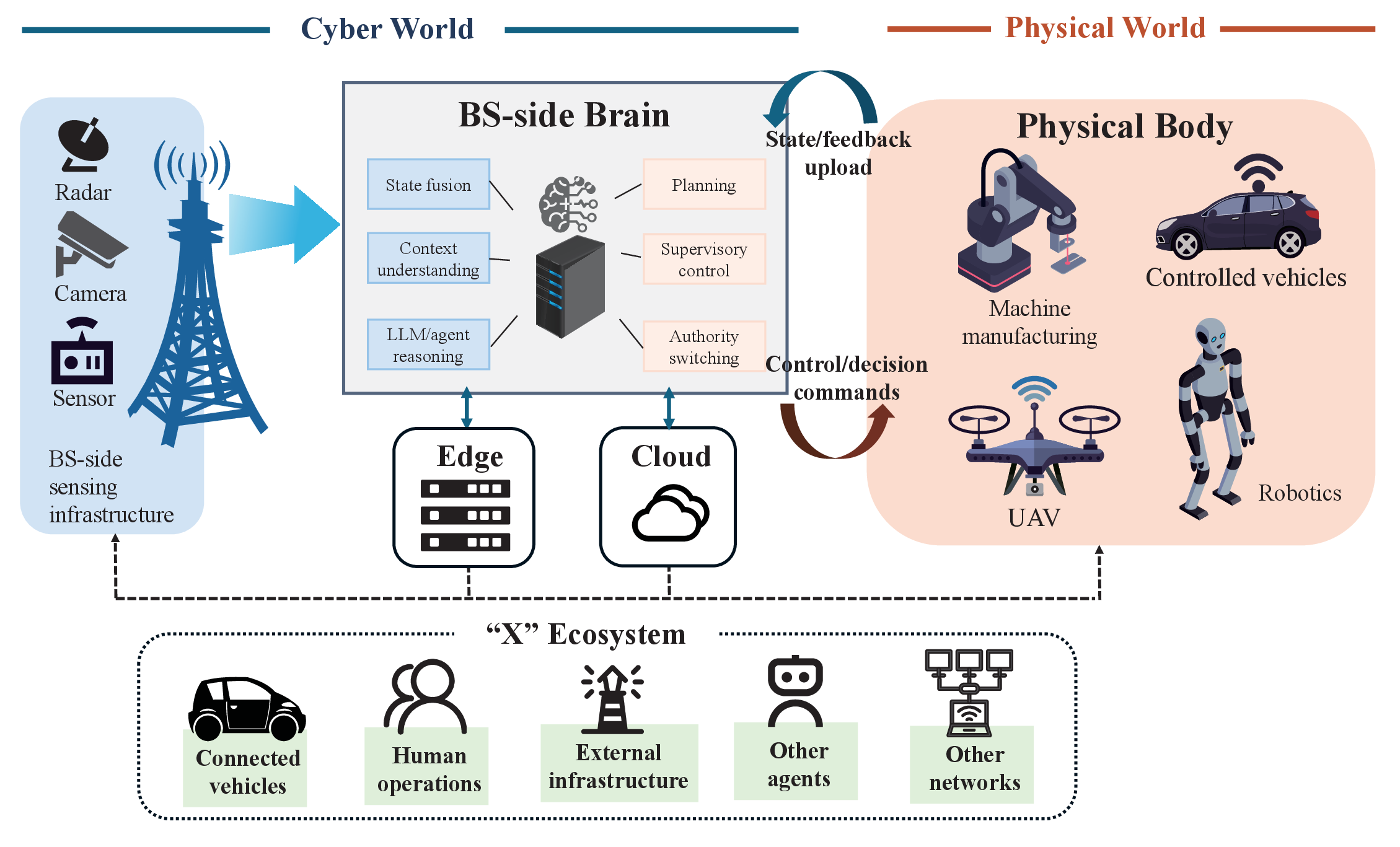}
\caption{Overall B2X framework of the physical world (Body) and cyber world (Brain), with edge/cloud as bridging layers for the ``X'' ecosystem.}
\label{fig:overall_framework}
\end{figure*}

Fig.~\ref{fig:overall_framework} represents the BS-side centralized B2X architecture as a closed-loop operating model rather than a static network diagram. The physical world contains embodied bodies such as industrial machines, connected vehicles, UAVs, and robots, while the cyber world contains the BS-side brain for state fusion, reasoning, planning, supervisory control, and authority switching with edge and cloud support. The uplink allows bodies to report physical states, sensory observations, and execution feedback to the brain. The downlink returns commands, task decisions, and coordination information to physical executors. Once this loop is established, wireless transmission becomes part of embodied operation itself. Specifically, the uplink design determines the timeliness and content of brain-side state knowledge, while the downlink design determines the reliability and regularity of decision delivery. This observation motivates the following sections on uplink state acquisition, downlink command delivery, and their joint communication-control design.

\section{Uplink B2X Networks}

This section studies B2X uplink transmission as the state-acquisition interface to the BS-side brain. In the BS-side centralized architecture, the uplink converts physical observations, task events, and execution feedback into network-side state awareness for state estimation and decision preparation.

\subsection{State Acquisition over the Uplink}

In conventional uplink design, performance is usually evaluated in terms of data rate, queue status, access efficiency, and packet reliability. In B2X, an uplink update is valuable when it improves the BS-side brain's understanding of the embodied task. A compact state feature may be more useful than a high-rate raw sensing stream for timely decision making, while a small safety event may require higher access priority than a larger but less urgent packet. Uplink state acquisition therefore depends on information content, access timing, and control relevance. The body may upload raw sensory data, compressed features, task events, local decisions, or control feedback, with different sizes, semantic values, interpretation latencies, and reliance on local autonomy. Updates should follow urgency, freshness, and physical dynamics. Reliability should be judged by its impact on state estimation, decision quality, and execution safety because delay, loss, and quantization errors affect BS-side inference. The key question is not how to maximize the aggregated upload volume, but which information refreshes network-side task understanding before it becomes stale.

\subsection{Uplink State-Reporting Regimes}

\begin{figure}[t]
    \centering
    \includegraphics[width=1\linewidth]{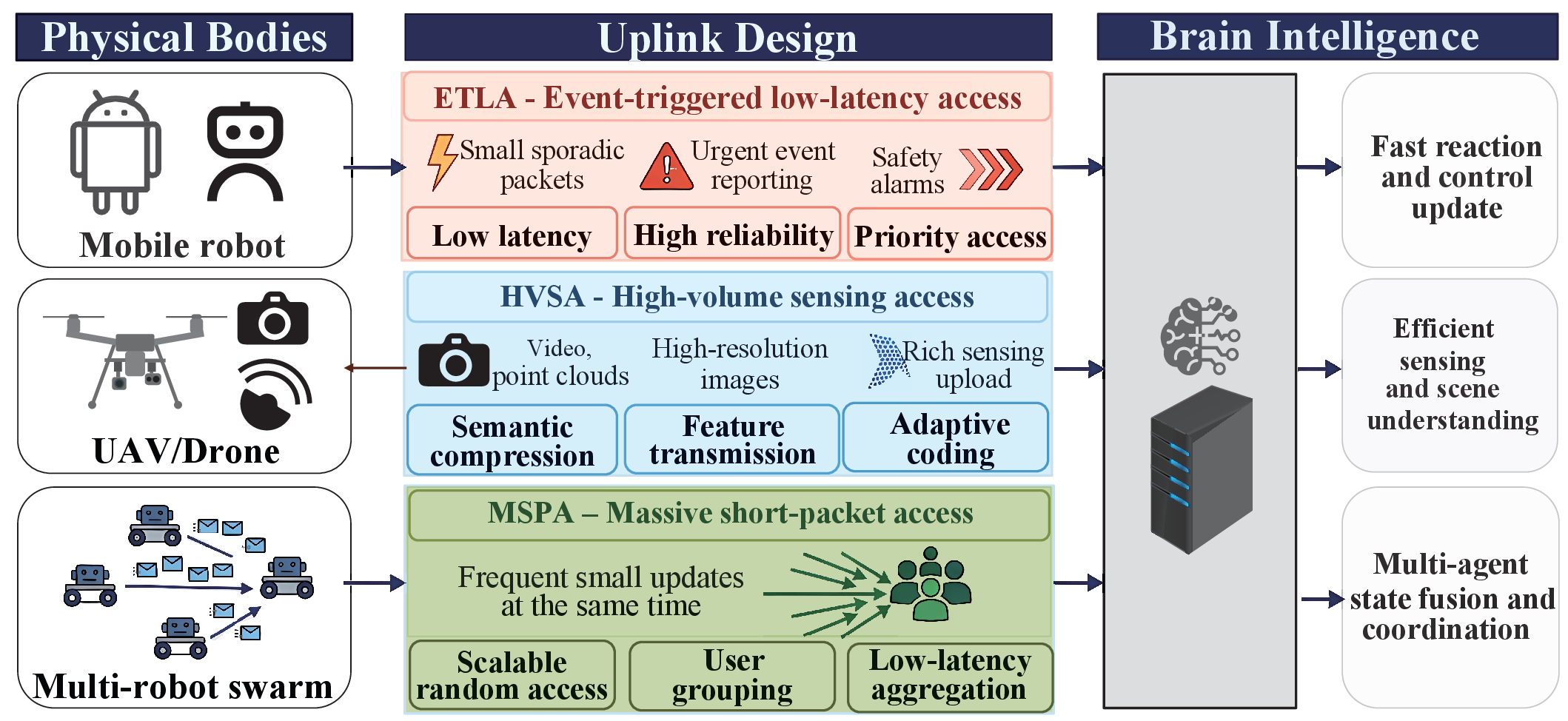}
    \caption{B2X uplink state-acquisition regimes for event-driven, sensing-intensive, and multi-body reporting.}
    \label{fig:uplink_design}
\end{figure}

Fig.~\ref{fig:uplink_design} summarizes three uplink state-reporting regimes, namely event-triggered low-latency access, high-volume sensing access, and massive short-packet access. These regimes correspond to event urgency, sensing volume, and simultaneous multi-body reporting, respectively, and reveal trade-offs among reaction latency, task-relevant fidelity, and coordination timeliness.

\begin{itemize}
    \item \textbf{Event-triggered low-latency access (ETLA):} ETLA addresses abrupt task-state changes that must reach the BS-side brain within the current control window. The payload is small, but its value for control decays quickly. The design priority is low reaction latency for rare but critical events without excessive resource reservation in normal operation. Mechanisms include event-triggered reporting, change-aware sampling, grant-free or configured-grant priority access, and short-packet transmission.
    \item \textbf{High-volume sensing access (HVSA):} HVSA arises when embodied devices serve as distributed sensing sources for the BS-side brain. UAVs, vehicles, robots, or intelligent sensors may upload images, video, point clouds, or multimodal features for network-side scene understanding and planning. The main trade-off is between sensing volume and task-relevant fidelity. Task-oriented compression, semantic feature transmission, unequal protection, edge-assisted reconstruction, and split inference reduce communication burden while preserving decision-relevant information.
    \item \textbf{Massive short-packet access (MSPA):} MSPA appears when many bodies report short distributed updates within a short time interval, as in multi-agent cooperation and distributed automation. Each packet may be short, but collisions, queueing, and uncoordinated access can prevent a timely BS-side system view. The design priority is scalable short-packet reporting with coordination timeliness. Mechanisms include scalable random access, user grouping, grant-light transmission, over-the-air aggregation, and age-aware scheduling.
\end{itemize}

ETLA, HVSA, and MSPA represent three ways in which uploaded information can become control-relevant: it can be urgent, information-rich, or collectively informative. The uplink therefore determines BS-side state awareness in terms of information content, freshness, and reliability for subsequent control.

\section{Downlink B2X Networks}

This section studies B2X downlink transmission as the command-delivery interface from the BS-side brain to embodied executors. Once the uplink has established state awareness, the downlink converts network-side decisions into executable instructions for control, coordination, and safety.

\subsection{Command Execution over the Downlink}

In the BS-side centralized B2X architecture, downlink transmission carries the information that closes the embodied loop, including low-level control commands, task-level decisions, coordination messages, model updates, and safety instructions. The importance of a downlink packet depends on its execution role, not only on data rate or delivery success. A delayed entertainment packet may reduce user experience, whereas a delayed motion command may alter the trajectory of an embodied device. Command execution therefore depends on command granularity, command cadence, and execution robustness. The BS-side brain may deliver direct actuator commands, trajectory segments, task decisions, or policy updates, with different sizes, interpretation latencies, autonomy levels, and body-side execution authority. Some tasks require periodic or semi-periodic updates for smooth control, while others need bursty commands when the task state changes, making command timing important for stability, responsiveness, and resource cost. Mobility, blockage, interference, and beam misalignment can weaken the downlink channel when reliable guidance is needed, so robustness depends on command executability under wireless uncertainty. The key issue is command delivery that preserves intended physical behavior, not packet arrival alone.

\subsection{Downlink Command-Delivery Regimes}

\begin{figure*}[t]
    \centering
    \includegraphics[width=0.8\linewidth]{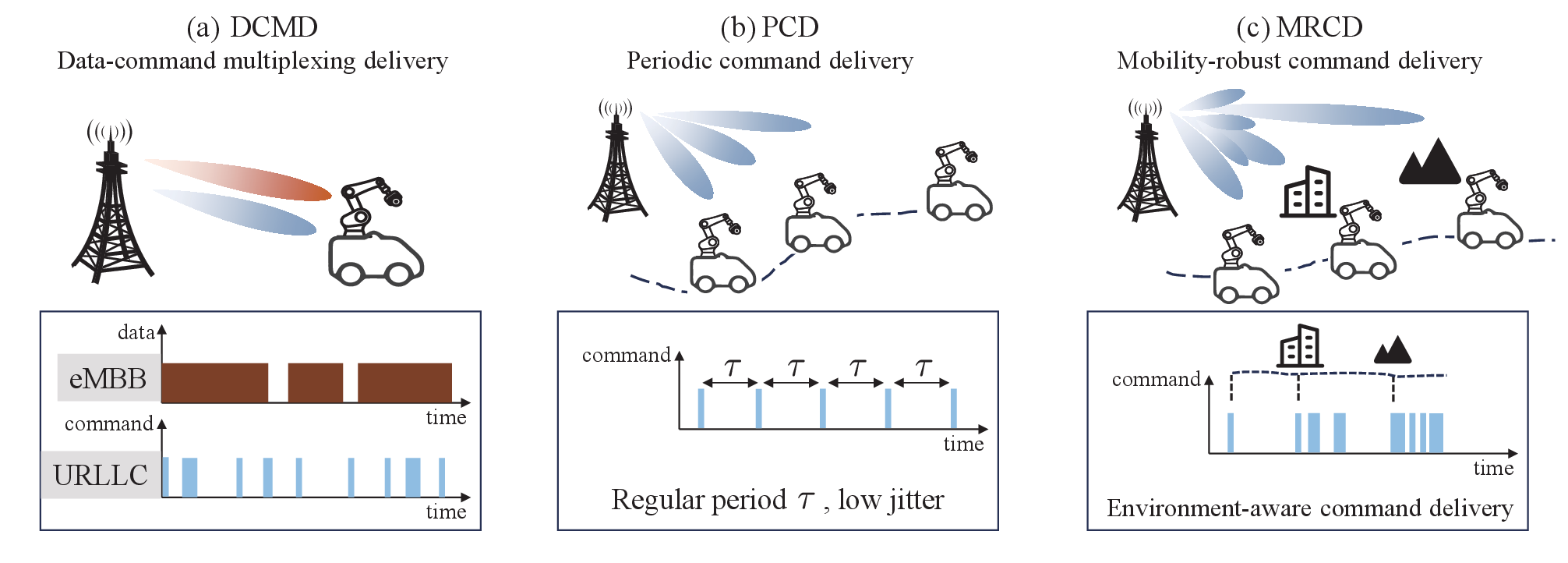}
    \caption{B2X downlink command-delivery regimes under shared BS resources.}
    \label{fig:b2x_downlink_trade-off}
\end{figure*}

Fig.~\ref{fig:b2x_downlink_trade-off}(a) illustrates a representative downlink setting in which a BS serves a communication user with long data packets and a control device with short command packets, corresponding to enhanced mobile broadband (eMBB) and ultra-reliable low-latency communications (URLLC) requirements~\cite{khan2022urllc}. It motivates three downlink command-delivery regimes, namely data-command multiplexed delivery, periodic command delivery, and mobility-robust command delivery, as illustrated in Fig.~\ref{fig:b2x_downlink_trade-off}.

\begin{itemize}
    \item \textbf{Data-command multiplexed delivery (DCMD):} DCMD arises when conventional service packets and control-command packets are transmitted through the same BS resource pool. Its distinguishing feature is traffic heterogeneity, since long data packets favor spectral efficiency and opportunistic scheduling, whereas short command packets require timely protected transmission. Downlink protocol design should coordinate packet segmentation, scheduling priority, and resource reservation so that service throughput does not suppress command urgency.
    \item \textbf{Periodic command delivery (PCD):} PCD addresses embodied tasks that rely on a continuing stream of command updates. The key requirement is low per-packet delay with regular inter-arrival times across successive commands. Compact command formats, configured transmission opportunities, and control-aware scheduling can reduce jitter and support smooth closed-loop execution.
    \item \textbf{Mobility-robust command delivery (MRCD):} MRCD appears when embodied executors move through changing channels or operate under blockage and interference. The central issue is command reliability as wireless channels and physical states vary together. Robust beam management, diversity support, priority protection, and control-aware power allocation can maintain command authority without assigning the same reliability level to all downlink traffic.
\end{itemize}

DCMD, PCD, and MRCD illustrate three ways in which delivered information becomes execution-critical. It can compete with service traffic, require regular command timing, or depend on robust delivery under mobility and channel uncertainty. The downlink therefore determines the decisions reaching the body, their arrival regularity, and their reliability for physical action. Uplink state acquisition and downlink command delivery jointly define the closed-loop communication-control trade-off studied next.

\section{Communication-Control Trade-off in B2X Networks}

This section integrates uplink and downlink design to enable a loop-level communication-control trade-off in B2X networks. At this level, state acquisition and command delivery are evaluated by their joint contribution to state freshness, command reliability, and control accuracy. This provides the basis for the Pareto characterization of B2X loop operation.

\subsection{From Link Optimization to Loop Optimization}

The previous two sections considered uplink and downlink design as separate wireless problems. This separation characterizes their individual roles, but it is not sufficient for B2X operation. The uplink determines BS-side acquisition of state and feedback information, while the downlink determines delivery of network-side decisions to the body for physical execution. The relevant optimization object is therefore the closed embodied loop formed by state acquisition, BS-side intelligence, command delivery, and physical actuation, rather than either wireless direction alone~\cite{lei2025edgehub}.

In this loop, the body generates physical states and feedback through sensing and interaction with the environment. The uplink reports this information to the BS-side brain for state fusion, reasoning, and planning. The downlink then delivers the resulting commands to the body for actuation, and the executed action leads to the next physical state. Link design affects B2X performance through this loop because communication resources ultimately support state freshness, command delivery, and closed-loop control quality.
\begin{itemize}
    \item \textbf{State freshness versus communication overhead.} The BS-side brain needs timely state and feedback information to generate control-relevant decisions. More frequent reporting, stronger packet error protection, and lower-latency access improve state freshness but also increase bandwidth use, access contention, and scheduling overhead. Uplink design should therefore be judged by loop-performance contribution, not only by per-report delay or reliability.
    \item \textbf{Command reliability versus radio resource use.} Downlink commands need to reach the body with sufficient timeliness and reliability to support physical execution. Stronger protection, regular command updates, and dedicated scheduling improve actuation support but also consume resources for communication traffic. The design question is whether the resource reservation is justified by command impact on the embodied loop.
    \item \textbf{State acquisition versus command delivery.} Uplink and downlink resources contribute to different parts of the same loop. More uplink resources improve BS-side state information, while more downlink resources improve command delivery. Since both directions share the same wireless budget, loop optimization should coordinate them according to their impact on joint control.
\end{itemize}

These trade-offs show that B2X optimization should be formulated at the loop level. The design objective is not independent uplink or downlink maximization, but the selection of communication and control operating points that jointly support closed-loop embodied operation. This motivates a Pareto characterization of B2X loop operation.

\subsection{Pareto Characterization of B2X Loop Operation}

Consider the BS-side centralized B2X architecture for a representative trajectory-control task, such as UAV inspection or robotic manipulation. The body senses its physical state and reports task-relevant observations and execution feedback to the BS-side brain. The BS-side brain fuses the reported state, computes a control decision, and sends the command over a beamformed wireless link. The loop objective is to maintain accurate trajectory execution while delivering control-relevant information with sufficiently small delay under shared radio resources.

Following the JDCC model in~\cite{gan2026modeling}, each loop operating point is described by two quantities: uplink information-delivery delay and normalized steady-state task-level control error. The delay reflects how quickly task-related state information reaches the BS-side brain, while the control error reflects the residual execution error after wireless-supported control is applied. In the above examples, this error corresponds to trajectory-tracking deviation for a UAV and to end-effector pose error for a robotic arm, which may lead to a missed grasp. Lower delay improves BS-side state freshness, whereas lower control error improves physical execution accuracy. These two improvements generally compete for radio and control resources, so each feasible communication-control policy corresponds to one point in the delay-error plane.

\begin{figure}[t!]
  \centering
  \includegraphics[width=0.9\linewidth]{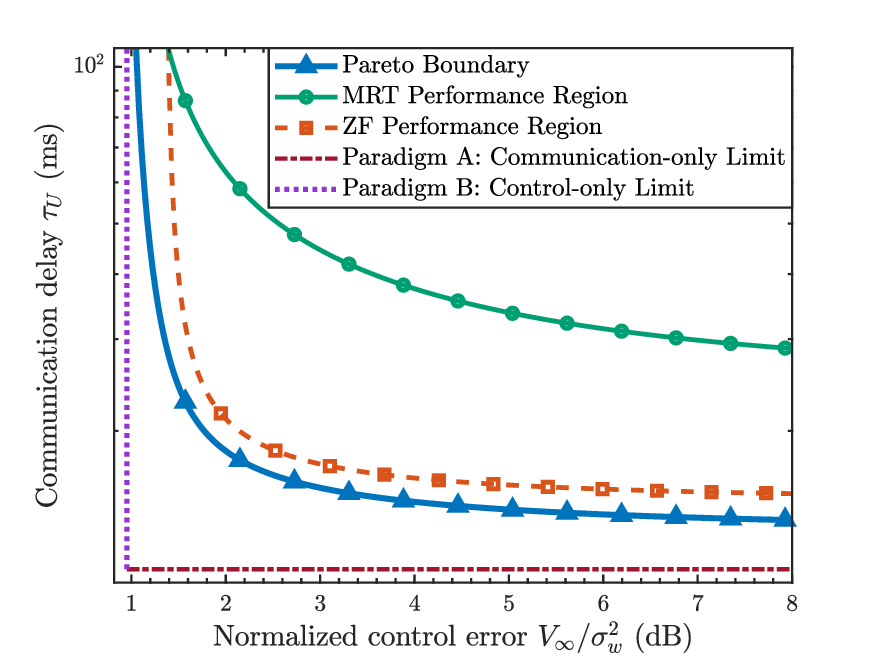}
  \caption{Delay-error Pareto characterization for the BS-side centralized B2X control loop, where $V_\infty/\sigma_w^2$ denotes normalized steady-state task execution error~\cite{gan2026modeling}.}
  \label{fig:pareto_boundary}
\end{figure}

Fig.~\ref{fig:pareto_boundary} illustrates the achievable delay-error region and its Pareto boundary for the considered B2X loop. The boundary summarizes the nondominated operating points, while the MRT and ZF curves serve as beamforming baselines with constrained transmission structures. The main implication is that B2X protocol and resource design should select a loop operating point according to the task's tolerance for delay and control error, rather than optimize a communication metric in isolation.

\section{Conclusion}

This article presented B2X networks as a wireless architecture that connects network-side intelligence with embodied execution. The distributed and centralized brain architectures specified intelligence placement, while the BS-side brain setting exposed uplink state acquisition and downlink command delivery as two transmission functions that create communication-control trade-offs. The Pareto-boundary case showed that communication efficiency and control quality form a coupled operating frontier rather than two independent objectives. Several research directions need further investigation for practical B2X design and deployment.

\begin{itemize}
    \item \textbf{Loop-level performance modeling and theory.} B2X needs a unified performance characterization that links wireless delay, reliability, sensing uncertainty, and feedback quality to tracking accuracy, stability, and safety. New analytical tools are needed to define meaningful operating points and identify whether system limits arise from communication, control, or their coupling.
    \item \textbf{Distributed brain placement and deployment.} Intelligence functions may be placed in the body-side, BS-side, or core-network-side brain, while edge and cloud resources provide additional computing support when needed~\cite{du2024distributed}. A key challenge is to place and adapt these functions according to task urgency, model complexity, reliability demand, and system load.
    \item \textbf{Function- and energy-aware protocol and resource co-design.} State reporting, control signaling, sensing-data sharing, and conventional traffic compete for shared radio, computing, and battery resources. In battery-constrained missions, such as UAV bridge inspection or continuous LiDAR mapping, B2X should jointly design access, packetization, scheduling, beamforming, and offloading to improve task performance per unit energy across sensing, computation, transmission, and actuation.
\end{itemize}

Progress in these directions will help B2X extend future wireless networks from communication infrastructure to intelligent platforms for connected vehicles, UAVs, robots, and other embodied agents. Its realization will require sustained progress in architecture, protocols, theory, and large-scale deployment, with JDCC serving as a core design principle throughout this evolution.

\vspace{5mm}
\noindent\textbf{Yuanwei Liu} (Fellow, IEEE) is a Professor at The University of Hong Kong and a visiting professor with Queen Mary University of London.\par\medskip

\noindent\textbf{Xu Gan} (Member, IEEE) is a Postdoctoral Fellow at The University of Hong Kong.\par\medskip

\noindent\textbf{Zhaolin Wang} (Member, IEEE) is a Research Assistant Professor at The University of Hong Kong.\par\medskip

\noindent\textbf{Chongjun Ouyang} (Member, IEEE) is a Postdoctoral Researcher at Queen Mary University of London.\par\medskip

\noindent\textbf{Hao Jiang} (Graduate Student Member, IEEE) is a Ph.D. student at Queen Mary University of London.\par\medskip

\noindent\textbf{Zongyao Zhao} (Member, IEEE) is a Postdoctoral Fellow at The University of Hong Kong.\par\medskip

\noindent\textbf{Kaibin Huang} (Fellow, IEEE) is a Professor at The University of Hong Kong.\par\medskip

\noindent\textbf{Robert Schober} (Fellow, IEEE) is an Alexander von Humboldt Professor and Chair for Digital Communication at FAU, Germany.

\end{document}